\documentclass[a4paper,10pt,twoside]{cpc-hepnp}
\usepackage{multicol}
\usepackage{graphicx}
\usepackage{amssymb,bm,mathrsfs,bbm,amscd}
\usepackage[tbtags]{amsmath}
\usepackage{lastpage}
\usepackage{CJK}
\usepackage{epstopdf}
\usepackage{epsfig}
\usepackage{color,xcolor}
\usepackage{ifpdf}
\usepackage{booktabs}

\usepackage{multirow}

\begin{document}

\fancyhead[c]{\small Chinese Physics C~~~Vol. $47$, No. 4 ($2023$) 044102} \fancyfoot[C]{\small 044102-\thepage}
\footnotetext[0]{}
\begin{CJK}{GBK}{song}
\title{Unified mechanism behind the even-parity ground state and neutron halo of $^{11}$Be \thanks{This work was partly supported by the Strategic Priority Research Program of Chinese Academy of Sciences (Grant No. XDB34000000), the National Natural Science Foundation of China under Grant No. 12275111 and the Fundamental Research Funds for the Central Universities (lzujbky-2021--sp41 and lzujbky-2021--sp36). }}
\author{%
      Jing Geng$^{1,2)}$ %
\quad Yi Fei Niu$^{1,2,3)}$ 
\quad Wen Hui Long$^{1,2,3)}$\email{longwh@lzu.edu.cn}  
}
\maketitle
\address{%
 $^1$Frontier Science Center for Rare isotope, Lanzhou University, Lanzhou 730000, China\\
 $^2$School of Nuclear Science and Technology, Lanzhou University, Lanzhou 730000, China\\
 $^3$Joint Department for Nuclear Physics, Lanzhou University and Institute of Modern Physics, CAS, Lanzhou 730000, China
}

\begin{abstract}
  Using the axially deformed relativistic Hartree-Fock-Bogoliubov (D-RHFB) model, we explore the mechanism behind the parity inversion and halo occurrence in $^{11}$Be, which are well reproduced by the RHF Lagrangian PKA1. It is illustrated that evidently enhanced deformation effects by the $\pi$-pseudo-vector and $\rho$-tensor couplings in PKA1 are crucial for correctly describing both even-parity ground state (GS) and neutron halo of $^{11}$Be. Coupling with the deformation, the intrude $1d_{5/2}$ component largely enhances the couplings between the even-parity orbit $1/2_2^+$ and the nuclear core to promise the even-parity GS, whereas the $2s_{1/2}$ component therein dominates the halo formation in $^{11}$Be. Moreover, the deformed halo in $^{11}$Be is found to be stabilized by the attractive inherent correlations between the $1d_{5/2}$ and $2s_{1/2}$ components of the halo orbit $1/2_2^+$, instead of pairing correlations, which paves a new way to understand the halo pictures in deformed unstable nuclei.
\end{abstract}

\footnotetext[0]{\hspace*{-3mm}\raisebox{0.3ex}{$\scriptstyle\copyright$}2013
Chinese Physical Society and the Institute of High Energy Physics
of the Chinese Academy of Sciences and the Institute
of Modern Physics of the Chinese Academy of Sciences and IOP Publishing Ltd}%

\end{CJK}

\begin{multicols}{2}

\section{Introduction}\label{Introduction}

Unstable nuclei far from the stability line in nuclear chart exhibit plenty of novel phenomena, for instance the emergence of new magic shells and the disappearance of the conventional ones \cite{Hoffman2008PRL100.152502, Simon1999PRL83.496, Motobayashi1995PLB346.9, TshooPRL109.022501, Ozawa2000PRL84.5493, Kanungo2009PRL102.152501, Steppenbeck2013Nature502.207}, the island of inversion \cite{Lenzi2010PRC82.054301, Crawford2019PRL122.052501, Ahn2019PRL123.212501}, and the dilute matter distributions -- halo structures \cite{Tanihata2013PPNP68.215, Minamisono1992PRL69.2058, Tanihata1985PRL55.2676}, which are attracting the interests of many researchers
in the fields of nuclear physics and related disciplines. While enriching the knowledge of nuclear physics, the emergence of novelties in unstable nuclei has also subverted our traditional understanding of atomic nuclei.

Beryllium (Be) isotopes are perfect candidates for exploring the novelty appearing in unstable nuclei. The novel parity inversion of $^{11}$Be is one of the significant issues. As deduced from the conventional shell model, the ground state (GS) of $^{11}$Be must have odd-parity because the odd neutron should populate the lower $p_{1/2}$ orbit. However, an early experiment with the reaction $^{11}$B($n,p$)$^{11}$Be indicated that $J^\pi=1/2^-$ is not the GS \cite{Wilkinson1959PR113.563}, which was further proofed to be even-parity $1/2^+$ by later experiments \cite{Donovan1961PR123.589, Alburger1964PR136.B916, Deutsch1968PL28B.178}. At a rather early stage, it was suggested that the even- and odd-parity states $1/2^\pm$ of $^{11}$Be should be very close \cite{Talmi1960PRL.4.469}. After a long time of silence, there were a number of attempts to interpret the GS parity of $^{11}$Be under the particle-core coupling picture \cite{Otsuka1993PRL70.1385, Sagawa1993PLB309.1, Esbensen1995PRC51.1274, Mau1995NPA592.33, Nunes1996NPA596.171, Gori2004PRC69.041302}, indicating that the quadruple core excitation and pairing blocking can be essential \cite{Sagawa1993PLB309.1, Esbensen1995PRC51.1274}. Besides, the deformation \cite{Ragnarsson1981NPA361.1, Li1996PRC54.1617, Pei2006NPA765.29}, the continuum \cite{Calci2016PRL117.242501} and effects beyond mean field \cite{Vinhmau1995NPA592.33, Bhattacharya1997PRC56.212} might also be also significant for correctly giving the even-parity GS of $^{11}$Be.

Despite the ingenious GS, special interests were also devoted to the halo occurrence in $^{11}$Be. Experimentally, $^{11}$Be has been confirmed to be of one-neutron halo structure \cite{Fukuda1991PLB268.339, Aumann2000PRL84.35, Schmitt2012PRL108.192701}. Theoretically, the energy density functional (EDF) theory presents dilute matter distribution for $^{11}$Be, while it fails to reproduce the even-parity GS \cite{Li1996PRC54.1617, Pei2006NPA765.29, Zhu1994PLB328.1}. Based on the variational shell model, it was indicated that the weak couplings between the core and odd neutron can be responsible for the one-neutron halo in $^{11}$Be \cite{Otsuka1993PRL70.1385}. Recently, the halo effects of $^{11}$Be on the one-neutron knockout reactions were studied using the halo effective filed theory \cite{HebbornPRC100.054607, HebbornPRC104.024616}. As a consensus, nuclear halo structures are contributed by neutrons or protons populating the weakly bound low-$l$ states, and the neighboring high-$l$ orbits do not contribute to the halo but stabilize the dilute nucleons via the pairing correlations \cite{Meng2006PPNP57.470}.

However, up to now both novelties in $^{11}$Be, i.e., the parity inversion and neutron halo, still cannot be interpreted in a unified framework, and the halo occurrence may essentially rely on the even-parity GS. The EDF theory, such as that adopted by the widely used Skyrme Hartree-Fock models, cannot uniformly describe the even-parity GS and halo occurrence in $^{11}$Be, as aforementioned. The relativistic mean field (RMF) theory \cite{Serot1986ANP16.1, Reinhard1989RPP52.439, Ring1996PPNP37.193, Bender2003RMP75.121, Meng2006PPNP57.470, Niksic2011PPNP66.519}, that stands on the microscopic meson exchange picture of the nuclear force, succeeds in describing the halo structures of spherical and deformed nuclei \cite{Meng2006PPNP57.470, Zhou2010PRC82.011301}, but fails to reproduce the parity inversion of $^{11}$Be. Implemented with the Fock diagrams, the relativistic Hartree-Fock (RHF) theory \cite{Bouyssy1987PRC36.380, Bernardos1993PRC48.2665, Marcos2004JPG30.703, Long2006PLB640.150, Long2007PRC76.034314} and an extension --- the relativistic Hartree-Fock-Bogoliubov (RHFB) theory \cite{Long2010PRC81.024308, Ebran2011PRC83.064323} have achieved many successes in describing various nuclear phenomena \cite{Long2022CTP74.097301}, including those of unstable nuclei \cite{Long2010PRC81.031302, Lu2013PRC87.034311, Li2016PRC93.054312, Li2019PLB788.192} and superheavy ones \cite{Li2014PLB732.169}.

In particular, the degrees of freedom associated with the $\pi$-pseudo-vector ($\pi$-PV) and $\rho$-tensor ($\rho$-T) couplings, important ingredients of the effective nuclear force, play significant roles in providing self-consistent descriptions of nuclear shell evolution \cite{Long2008EPL82.12001, Long2009PLB680.428, Wang2013PRC87.047301}, the pseudo-spin symmetry restoration \cite{Long2007PRC76.034314, Geng2019PRC100.051301R} and the emergences of new magicity \cite{Li2016PLB753.97, Liu2020PLB806.135524}. Recently, considering nuclear deformation that may be significant for $^{11}$Be, the axially deformed RHF \cite{Geng2020PRC101.064302} and RHFB \cite{Geng2022PRC105.034329} models, respectively called the D-RHF and D-RHFB models, have been established by utilizing an expansion of the spherical Dirac Woods-Saxon (DWS) base \cite{Zhou2003PRC68.034323}. It was revealed that the $\pi$-PV and $\rho$-T couplings, which contribute mainly via the Fock diagrams, evidently enhance the deformation effects of nuclei \cite{Geng2020PRC101.064302, Geng2022PRC105.034329}.

In particular, the D-RHFB model has the advantage of a unified treatment of spin-orbit coupling, tensor force, deformation, pairing correlations and the continuum \cite{Geng2022PRC105.034329}, the significant mechanisms which are essential for unstable nuclei. This encourages us to explore the underlying mechanism related to the even-parity GS and halo occurrence in $^{11}$Be. The paper is organized as follows. In Sec. {\ref{sec:General Formalism}}, the decomposition of the canonical single-particle energy and its physical significance are introduced. Afterwards, the parity inversion and the halo phenomenon are discussed in Sec. {\ref{sec:Results}}. Finally, we give a summary in Sec. {\ref{sec:Summary}}.

\section{Canonical s.p. energy and the decomposition}\label{sec:General Formalism}

To understand the microscopic picture of $^{11}$Be, the language of the basis expansion is rather helpful. In the D-RHFB model, the spherical DWS base is used to expand the quasi-particle wave functions $\psi^V$ and $\psi^U$ \cite{Geng2022PRC105.034329},
\begin{equation}\label{eq:PSI}
    \psi_{\nu\pi m}^V = \sum_{a} C_{\nu\pi m,a}^V \psi_{am}, \hspace{1em} \psi_{\nu\pi m}^U =  \sum_a C_{\nu\pi m, a}^U \psi_{am},
\end{equation}
where $m$ and $\pi$ denote the angular momentum projection and parity, $\nu$ stands for the index of the orbits in the $\pi m$ block. Imposing axial symmetry and reflection symmetry, $m$ and $\pi$ remain as good quantum numbers. For a spherical DWS base, $\psi_{am}$ represents the wave function and the index $a$ = $(n\kappa)$ carries the principle number $n$ and $\kappa$-quantity, namely $\kappa = \pm (j+1/2)$ with $j=l\mp1/2$, where $l$ is the orbital angular momentum. In the following, the index $i = (\nu\pi m)$ is used to denote the deformed orbits, while the index $a$ specifies the states in the spherical DWS base. It is worth noting that the DWS base has the advantage of providing appropriate asymptotic behaviors for the wave functions due to the nature of the Woods-Saxon type potential \cite{Woods1954PhysRev95.577}. For unstable nuclei, e.g. $^{11}$Be with a halo structure, this advantage is essential to provide a reliable description.

For the convenience of analysis, the canonical transformation from the quasi-particle space is performed in general to get canonical s.p. states, by diagonalizing the density matrix. In practice, the canonical s.p. space is taken to be similar to the Hartree-Fock space, for both of which the density matrices are diagonal. Utilizing the orthogonality of the DWS base, there is a straightforward way to build the density matrix, which reads as,
\begin{equation}
  \rho_{aa'}^{\pi m} =  \sum_{\nu} C_{\nu\pi m, a}^V C_{\nu\pi m, a'}^V.
\end{equation}
Notice that for the contributions from the odd-nucleon in the blocked orbit, the expansion coefficient $C^V$ shall be replaced by the $C^U$ one in Eq. (\ref{eq:PSI}). The diagonalization of the above density matrix $\rho_{aa'}^{\pi m}$ gives the eigenvalues $v_i^2$, namely the occupation probability of orbit $i = m_\nu^\pi$, and the eigenvectors $\widehat D_{\nu}^{\pi m}$, the set of the expansion coefficients $D_{i,n\kappa}$ of the canonical wave functions in the spherical DWS base.

Thus, in terms of the spherical DWS base, the canonical s.p. energy $\varepsilon_i$ and wave function $\psi_i$ for orbit $i$ can be expressed as,
\begin{equation}\label{eq:Eig}
  \varepsilon_i =  \sum_{aa'} D_{i,a} h_{aa'}^{\pi m}D_{i,a'}, \hspace{1em}
  \psi_{\nu\pi m} =  \sum_{n\kappa} D_{i,n\kappa} \psi_{n\kappa m},
\end{equation}
where $h_{aa'}$ is the s.p. Hamiltonian in the spherical DWS base, the index $a$ is for spherical DWS basis states. Following the deformation of a nucleus, orbits which originally had the same spherical $n\kappa$ quantities cannot degenerate any more, leading to a mixture of spherical $n\kappa$-components in deformed s.p. orbits. As seen from Eq. (\ref{eq:Eig}), the couplings between various spherical DWS components $a$ and $a'$ give the canonical s.p. energy. Thus, after summing the principle numbers $n$ and $n'$, thsy can be decomposed into the diagonal terms $E_{d,i}^\kappa$ and off-diagonal ones $E_{c,i}^{\kappa\kappa'}$ as,
\begin{equation}
  \varepsilon_i = \sum_{\kappa\kappa'} E_i^{\kappa\kappa'} =  \sum_\kappa p_{i\kappa}^2 E_{d,i}^\kappa + \sum_{\kappa\ne\kappa'} p_{i\kappa} p_{i\kappa'} E_{c,i}^{\kappa\kappa'},
\end{equation}
where $p_{i\kappa}^2$ describes the sum proportion of the $\kappa$-component of the orbit $m_\nu^\pi$, and $E_{i}^{\kappa\kappa'}$ corresponds to the coupling effects between the spherical $\kappa$ and $\kappa'$ components,
\begin{equation}
  p_{ik} = \sum_n D_{i,n\kappa}^2, \hspace{1em} E_{i}^{\kappa\kappa'} = \sum_{nn'} D_{i,n\kappa} h_{aa'}^{\pi m} D_{i,n'\kappa'}.
\end{equation}
Using the above conventions, the diagonal term $E_{d,i}^\kappa$ and the off-diagonal one $E_{c,i}^{\kappa\kappa'}$ can be defined as,
\begin{equation}\label{eq:Expansion-DC}
  E_{d,i}^\kappa \equiv  E_{i}^{\kappa\kappa}/p_{i\kappa}^2, \hspace{1em}E_{c,i}^{\kappa\kappa'} \equiv E_{i}^{\kappa\kappa'}/\big(p_{i\kappa} p_{i\kappa'}\big).
\end{equation}
Further considering the sum over $\kappa$ components, the diagonal contribution $E_{d,i}$ and the off-diagonal one $E_{c,i}$ to the canonical s.p. energy $\varepsilon_i$ read as,
\begin{equation}\label{eq:Expansion-DCi}
  E_{d,i} =  \sum_\kappa E_{i}^{\kappa\kappa}, \hspace{1em} E_{c,i} =  \sum_{\kappa\ne\kappa'} E_{i}^{\kappa\kappa'}.
\end{equation}

As aforementioned, more spreading of the proportion $p_{i\kappa}^2$ over different $\kappa$-components is in general due to enhanced deformation effects. Thus using the language of the spherical DWS basis expansion, the $E_{d,i}^\kappa$-term can be taken as the centroid energy of the $\kappa$-component of the deformed orbit $i$, or taken as the single-particle-like energy of the spherical $\kappa$-fragment of orbit $i$. In contrast, the $E_{c,i}^{\kappa\kappa'}$-term describes the inherent correlation between the $\kappa$- and $\kappa'$-components, which in fact reflects the shape evolution of orbit $i$. Following deformation, it is not hard to deduce that orbit $i$ becomes more bound with the negative $E_{c,i}$ value, and vice versa.

\section{Unified mechanism for parity inversion and neutron halo in $^{11}$Be}\label{sec:Results}

Aiming at a unified mechanism behind the even-parity GS and neutron halo of $^{11}$Be, we focus on the D-RHFB calculations with PKA1 \cite{Long2007PRC76.034314}, one of the most complete RHF Lagrangians which contains the $\pi$-PV and $\rho$-T couplings. For comparison, the RHF Lagrangians PKO3 \cite{Long2008EPL82.12001} containing $\pi$-PV coupling and PKO2 \cite{Long2008EPL82.12001}, and the RMF one DD-LZ1 \cite{Wei2020CPC44.074107} are considered as well. In the pairing channel, the finite-range Gogny force D1S \cite{Berger1984NPA428.23} is utilized as the pairing force, due to the advantage of a finite range, i.e., natural convergence with the configuration space. The details of the calculations can be found in Ref. \cite{Geng2022PRC105.034329}.

To find the global minimum for an odd-$A$ nucleus using the D-RHFB model, the blocking effects shall be treated carefully \cite{Ring1980Springer-Verlag}. Similar to a popular treatment in the RMF calculations, the equal filling approximation is adopted \cite{LI2012CPL29.42101,Perez-Martin2008PRC78.014304}. Within the Bogoliubov scheme, we attempted to block the first quasi-particle level for various $\pi m$-block to determine the ground state. According to the mapping relation between the canonical single-particle and the Bogoliubov quasi-particle states, the blocked quasi-particle orbits are mapped in general with the s.p. ones closest to the Fermi level. For $^{11}$Be, we tried to block the states $m^\pi=$ $1/2^-$, $1/2^+$, $3/2^+$ and $5/2^+$, and only show the results of $m^\pi = 1/2^-$ and $1/2^+$, from which the ground state can be determined by selected models.

\subsection{Parity inversion of $^{11}$Be}

We first performed the shape constrained calculations for $^{11}$Be using PKA1, PKO2, PKO3 and DD-LZ1. Figures \ref{Fig:Ebeta} (a) and \ref{Fig:Ebeta} (b) show the binding energy $E_B$ (MeV) of $^{11}$Be as a function of the quadruple deformation $\beta$, which corresponds to the blocking of even- and odd-parity states $m^\pi = 1/2^\pm$, respectively. It is shown that only PKA1 results in the same even-parity GS as the experiments \cite{Wilkinson1959PR113.563, Donovan1961PR123.589, Alburger1964PR136.B916}. Moreover, a rather large prolate deformation ($\beta\simeq1.2$) is obtained by PKA1 for the GS, while the others give an odd-parity GS with spherical or near spherical shapes. Compared with the experimental value, which reads as $-$65.478 MeV \cite{Wang2012CPC36.1603}, the binding energy given by PKA1 is overestimated by roughly $3$ MeV. In fact, the calculations of nearby nuclei show that the binding energies for light nuclei are overestimated by PKA1 in general.

\begin{center}
  \includegraphics[width=0.98\linewidth]{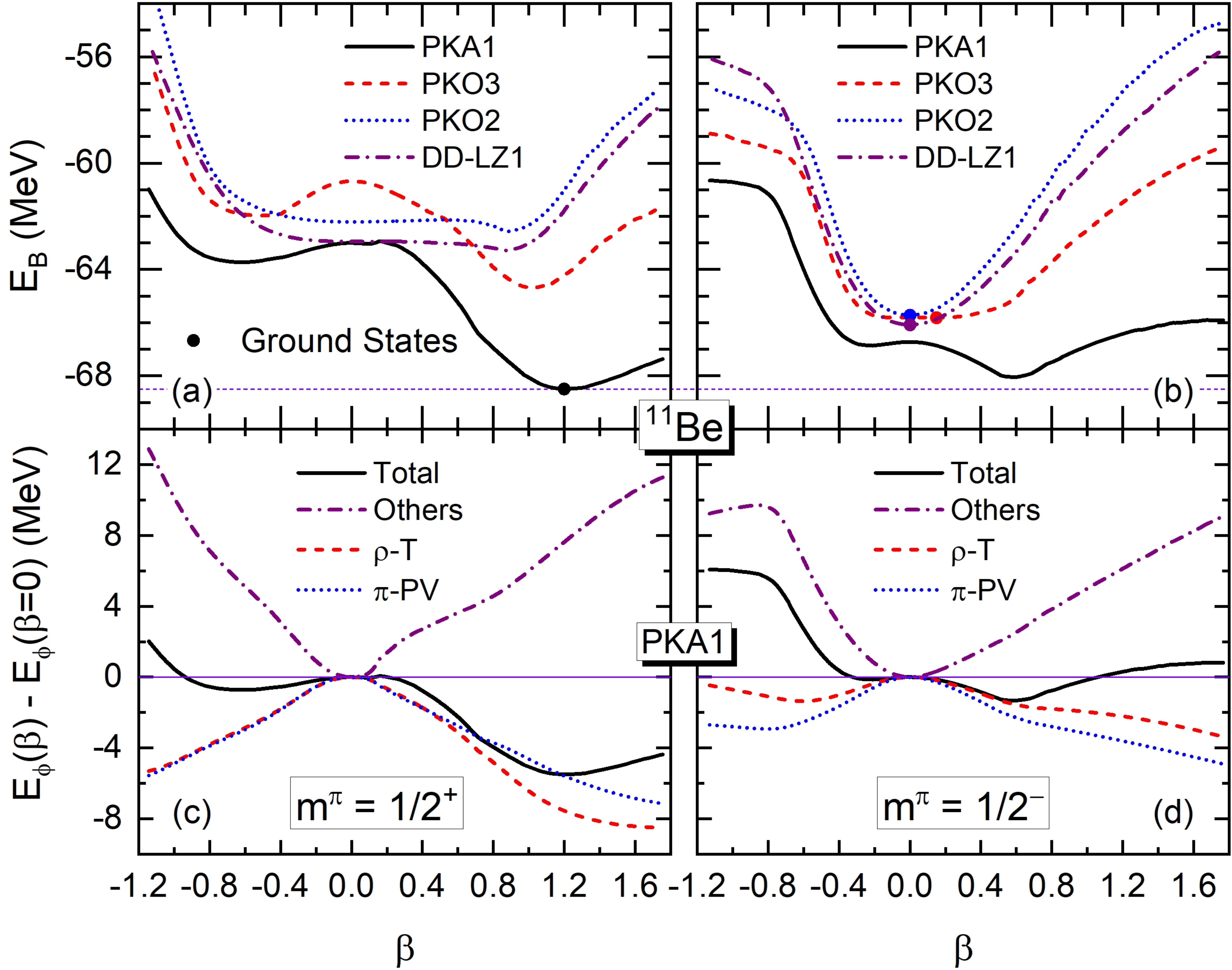}
  \figcaption{(color online) Binding energies $E_B$ (MeV) as functions of the quadruple deformation $\beta$ for $^{11}$Be with even [plot (a)] and odd [plot (b)] parity. The results are calculated by PKA1, PKO2, PKO3 and DD-LZ1. Referring to $\beta=0$, plots (c) and (d) show the PKA1 contributions of $E_B$ respectively with even and odd parity, including the total, the $\pi$-PV and $\rho$-T ones, and the sum from the other channels.}\label{Fig:Ebeta}
\end{center}

As an example, Table \ref{tab:bulk} shows the calculated binding energies $E_B$ of $^{10,11,12}$Be, as well as the one-neutron separation energy $S_{n}$ for $^{11}$Be and the two-neutron one $S_{2n}$ for $^{12}$Be, in comparison with the experimental data \cite{Wang2017CPC41.030003}. It can be found that the systematics of the binding energies, here from $^{10}$Be to $^{12}$Be, are different between PKA1 and the other selected models. Specifically, all the $E_B$ values given by PKA1 are larger than the experimental ones by $1.4\sim 3.5$ MeV. While the binding energy of $^{10}$Be is underestimated by all other selected Lagrangians, the one of $^{12}$Be is overestimated by different extents. However, for the $S_n$ and $S_{2n}$ values, all selected models provide a similar accurate result as the experimental values. Thus, deviations in the binding energy of a few MeV do not do not influence the conclusions made in this work substantially.

\begin{center}
\tabcaption{Binding energies $E_B$ (MeV) of $^{10,11,12}$Be, along with the one-neutron separation energy (MeV) $S_{n}$ for $^{11}$Be and two-neutron one $S_{2n}$ for $^{12}$Be calculated by PKA1, PKO2, PKO3 and DD-ME2, as compared to the experimental (Exp.) data \cite{Wang2017CPC41.030003}. } 

\label{tab:bulk}\renewcommand{\arraystretch}{1.5}\setlength{\tabcolsep}{0.5em}
\begin{tabular}{cccccc}  
\toprule[0.15em]
       &          \multicolumn{3}{c}{$E_B$}   &     $S_{n}$ &   $S_{2n}$ \\ \cline{2-4} \cline{5-5} \cline{6-6}
       &  $^{10}$Be &  $^{11}$Be &  $^{12}$Be  &   $^{11}$Be &  $^{12}$Be \\\hline
Exp.   &  $-$64.98  &   $-$65.48 &   $-$68.65  &        0.50 &       3.67 \\
PKA1   &  $-$66.40  &   $-$68.50 &   $-$72.21  &        2.10 &       5.81 \\
PKO3   &  $-$63.63  &   $-$65.82 &   $-$70.13  &        2.19 &       6.50 \\
PKO2   &  $-$63.50  &   $-$65.73 &   $-$69.00  &        2.23 &       5.50 \\
DD-LZ1 &  $-$63.61  &   $-$66.09 &   $-$69.67  &        2.47 &       6.06 \\
\bottomrule[0.15em]
\end{tabular}

\end{center}

In order to clarify the mechanism behind the GS of $^{11}$Be, Figs. \ref{Fig:Ebeta} (c) and \ref{Fig:Ebeta} (d) show the contributions of $E_B$ referring to $\beta=0$ given by PKA1 for even- and odd-parity cases, respectively, namely the total,  the $\pi$-PV one that is enclosed in both PKA1 and PKO3, the $\rho$-T one which is uniquely considered by PKA1, and the sum from other channels, which corresponds to all degrees of freedom adopted in PKO2 and DD-LZ1. It is interesting to see that both $\pi$-PV and $\rho$-T contributions are evidently enhanced following the enlarged deformation. Compared with the odd-parity one [plot (d)], more remarkable enhancements are presented by the $\pi$-PV and $\rho$-T couplings for the even-parity case [plot (c)], which are seen to be crucial for reproducing the even-parity GS of $^{11}$Be.

In fact, referring to other selected models that fail to reproduce the GS of $^{11}$Be, it is also helpful to understand the role played by the $\pi$-PV and $\rho$-T couplings. In Figs. \ref{Fig:Ebeta} (a) and \ref{Fig:Ebeta} (b), both PKO2 and DD-LZ1, which do not contain either $\pi$-PV or $\rho$-T couplings, present the odd-parity GS with a fairly stable spherical shape and even-parity local minima with a rather soft prolate shape. For PKO3, which contains the $\pi$-PV coupling, an odd-parity GS with a rather soft near-spherical shape and an even-parity local minimum with a solid prolate shape are obtained, and the energy difference between the even- and odd-parity minima are significantly reduced, as compared with PKO2 and DD-LZ1. Thus, combined with the PKA1 results, it can be concluded that both $\pi$-PV and $\rho$-T couplings, which evidently enhance the deformation effects, play a key role in determining the even-parity GS of $^{11}$Be.

\begin{figure*}
\begin{center}
  \includegraphics[width=0.5\linewidth]{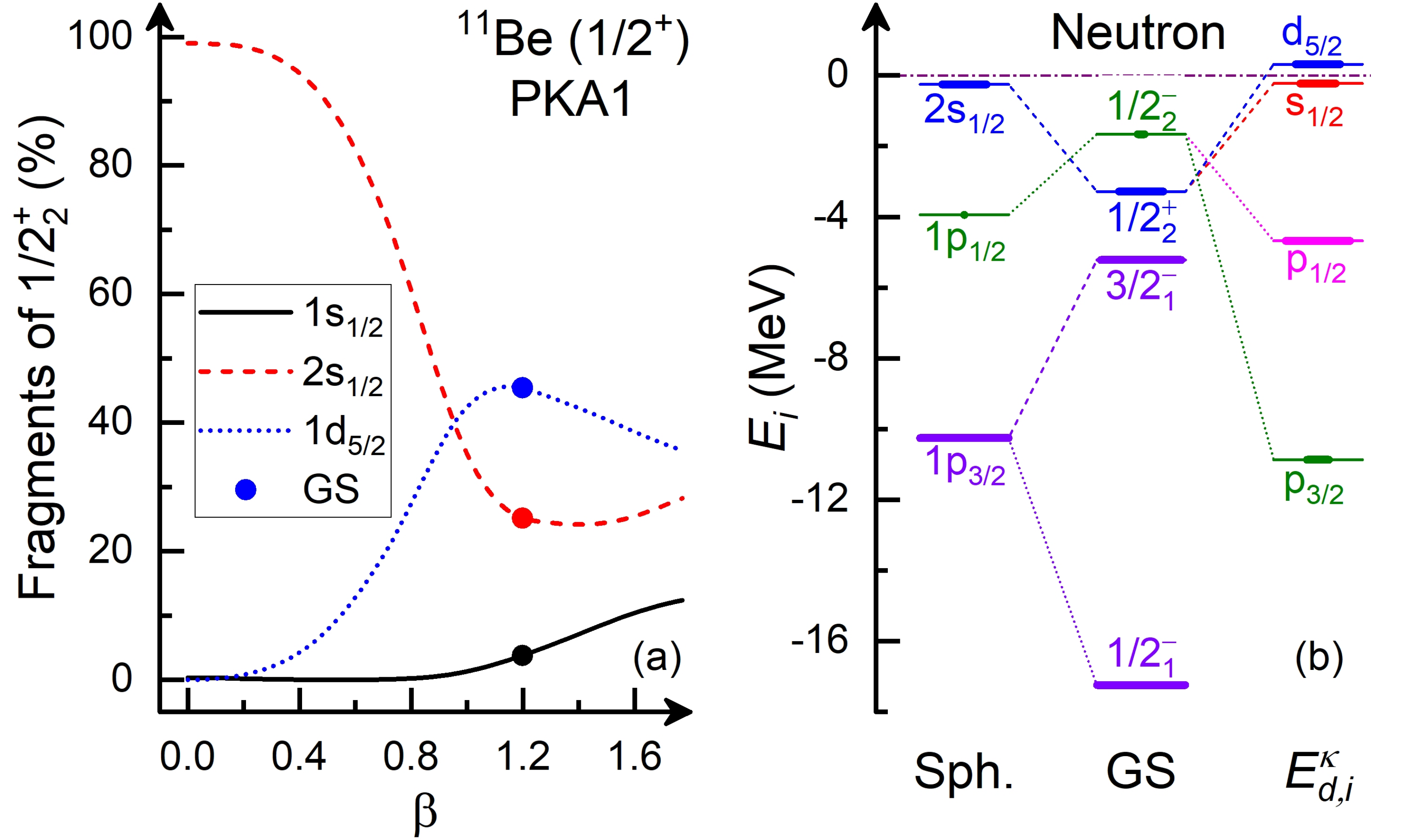}
  \figcaption{(color online) Expansion proportions of neutron orbit $1/2_2^+$ with respect to the deformation $\beta$ [plot(a)], and the canonical s.p. energies $E_i$ (MeV) and occupation probabilities (in thick bars) of spherical (Sph.) $^{11}$Be and the GS [plot (b)]. The results are extracted from the PKA1 calculations by blocking the state $m^\pi = 1/2^+$, and the $E_{d,i}^\kappa$ values (MeV) and expansion proportions (in thick bars) of the neutron orbits $1/2_2^\pm $ of the GS are also shown in plot (b). } \label{Fig:SPK-Be}
\end{center}
\end{figure*}

To understand why the even-parity GS of $^{11}$Be can be correctly reproduced at large deformation in terms of the spherical DWS base, the expansion proportions of the even-parity orbit $1/2_2^+$ occupied by the odd neutron are shown as functions of the deformation $\beta$ in Fig. \ref{Fig:SPK-Be} (a). Figure \ref{Fig:SPK-Be} (b) shows the canonical s.p. energies $E_i$ (MeV) with occupation probabilities (in thick bars) for spherical $^{11}$Be and the GS, and the $E_{d,i}^\kappa$ values deduced from Eq. (\ref{eq:Expansion-DC}) with the expansion proportions (in thick bars) of the dominant $\kappa$ fragments of the orbits $1/2_2^\pm$ for the GS. The results are extracted from the PKA1 calculations by blocking the state $m^\pi = 1/2^+$. As shown in Fig. \ref{Fig:SPK-Be} (a), the $1d_{5/2}$ component in the orbit $1/2_2^+$ increases gradually with a simultaneously reduced $2s_{1/2}$ component when $^{11}$Be becomes prolately deformed. Meanwhile, the inversion of the even- and odd-parity orbits $1/2_2^\pm$ from the spherical $^{11}$Be to the GS is found in Fig. \ref{Fig:SPK-Be} (b) and is consistent with an even-parity GS.

Moreover, as deduced from the high-lying $\kappa$ fragments of the $1/2_2^+$ orbit in Fig. \ref{Fig:SPK-Be} (b), the $E_{c,i}$ value given by Eq. (\ref{eq:Expansion-DCi}) shall be negatively and large, indicating strong attractive inherent correlations between the $\kappa$ fragments of the $1/2_2^+$ orbit. In contrast, the $\kappa$ fragments of the odd-parity orbit $1/2_2^-$ are deeply bound, and the deduced large, positively $E_{c,i}$ value reveals strong repulsive inherent correlations between the $\kappa$ fragments of the $1/2_2^-$ orbit. Thus, from the opposite effects carried by the correlation terms $E_{c,i}$, it is not hard to understand the inversion of the even- and odd-parity orbits $1/2_2^\pm$ from spherical $^{11}$Be to the deformed GS, in which the enhancements due to the $\rho$-T and $\pi$-PV couplings are very important as deduced from the nature of the channels \cite{Geng2022PRC105.034329}.

In fact, the PKA1 results are consistent with the indication by Bohr and Mottelson that for a $p$-shell nucleus $^{11}$Be the even-parity GS can be a consequence of a strong preference for the orbit $[220,1/2]$ with prolate deformation \cite{Bohr1998Singapore}. Supplementary, it should be emphasized that the expansion proportions in the theoretical calculation are different from the spectroscopic factor, which is a basic quantity characterizing the single-particle nature of nuclear excitations in both experiments and theories. In this work, the calculations of $^{11}$Be are performed by restrictions on the mean-field level. Thus, one cannot take the expansion proportions in Fig. \ref{Fig:SPK-Be} (a) as some equivalent concept of the spectroscopic factor, or even compare it directly with the measured data of the transfer reaction (see Ref. \cite{Schmitt2012PRL108.192701} and references there in).

\begin{center}
  \includegraphics[width=0.98\linewidth]{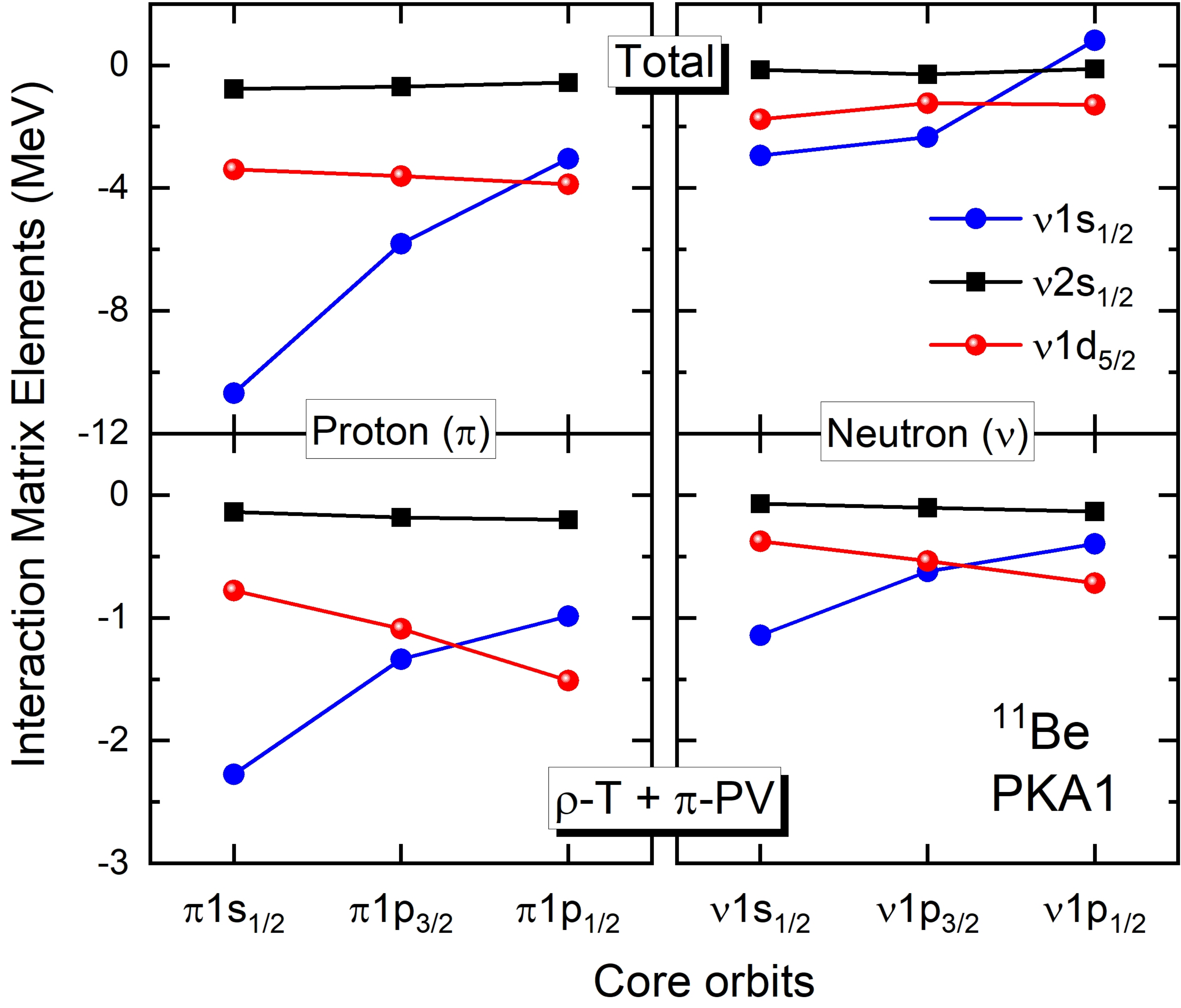}\\
  \figcaption{(color online) Interaction matrix elements (MeV) between a selected neutron ($\nu$) and the core orbits of $^{11}$Be. The left and right plots correspond to the couplings with proton ($\pi$) and neutron ($\nu$) cores, respectively, including the total (upper plots) and the sum of the $\rho$-T and $\pi$-PV couplings (lower plots). The results are obtained from spherical RHFB calculations with PKA1 by blocking neutron $s$ state.  }\label{Fig:IME-sp}
\end{center}

To enhance the understanding of the largely deformed GS and the role played by the $\pi$-PV and $\rho$-T couplings, Fig. \ref{Fig:IME-sp} shows the interaction matrix elements (IMEs) between selected neutron states and core orbits of $^{11}$Be, including the total (upper plots) and the sum of the $\rho$-T and $\pi$-PV couplings (lower plots). The results are given by the spherical RHFB calculations \cite{Long2010PRC81.024308} with PKA1 by blocking the even-parity state $s_{1/2}$. In fact, the selected even-parity states $\nu1s_{1/2}$, $\nu2s_{1/2}$ and $\nu1d_{5/2}$ are the dominant orbital components of the deformed orbit $1/2_2^+$, as seen from Fig. \ref{Fig:SPK-Be} (a).

As the general feature of the IMEs in Fig. \ref{Fig:IME-sp}, the proton $(\pi)$ core (left plots) shows stronger couplings with selected neutron orbits than the neutron $(\nu)$ core (right plots). This can be explained by the effects of the Fock terms \cite{Bouyssy1985PRL55.1731}. Specifically, the Fock terms arising from the isoscalar $\sigma$- and $\omega$-couplings do not contribute to neutron-proton ($np$) interactions but present net repulsive neutron-neutron ($nn$) and proton-proton ($pp$) ones, thus partly canceling their net attractive Hartree terms in the $nn$ and $pp$ channels. In contrast, as shown in the lower plots of Fig. \ref{Fig:IME-sp}, the isovector $\rho$-T and $\pi$-PV couplings present rather strong attractive contributions mainly through the Fock terms, particularly for the $np$ interactions (left-lower plot). Compared with other RHF and RMF models, the $np$ interactions are evidently enhanced due to the strong $\rho$-T coupling in PKA1 \cite{Long2010PRC81.031302, Geng2019PRC100.051301R}.

Combined with Fig. \ref{Fig:SPK-Be}, the IMEs in Fig. \ref{Fig:IME-sp} can help us understand the evident deformation effects given by PKA1 in Fig. \ref{Fig:Ebeta}. As seen from Fig. \ref{Fig:IME-sp}, the neutron state $\nu2s_{1/2}$, namely the $1/2_2^+$ orbit at $\beta=0$, shows rather weak couplings with both neutron and proton cores due to the nodal differences, and thus a loosely bound $2s_{1/2}$ orbit is found in Fig. \ref{Fig:SPK-Be} (b), which indicates an unfavorable even-parity configuration for spherical $^{11}$Be. Compared with $\nu2s_{1/2}$, the $\nu1d_{5/2}$ state shows much stronger couplings with the cores. Combined with the gradually enlarged $1d_{5/2}$ proportion and the simultaneously reduced $2s_{1/2}$ one in Fig. \ref{Fig:SPK-Be} (a), the couplings between the $1/2_2^+$ orbit and the cores are then notably enhanced following the enlarged deformation, which eventually leads to a fairly deep bound $1/2_2^+$ orbit in the GS of $^{11}$Be. In addition, as shown in Fig. \ref{Fig:SPK-Be} (a), the $1s_{1/2}$ component is also enfolded slightly into the orbit $1/2_2^+$ in the GS, which also plays some role in deepening the orbit $1/2_{2}^+$, due to its rather strong couplings with the core orbits in Fig. \ref{Fig:IME-sp}.

In contrast, for odd-parity $^{11}$Be, which is not shown in details, the mixing of the $1p_{1/2}$- and $1p_{3/2}$-components also happens in the populated odd-parity orbits $1/2_{1}^-$ and $1/2_2^-$ following an increase in the deformation. However, the mixing of high-lying odd-parity components such as $2p$ and $1f$ are hindered by the $sd$-shell between the $p$- and $pf$-shells. This indicates that the sums of the $1p_{3/2}$ and $1p_{1/2}$ proportions in these orbits will not change much, which is confirmed by the D-RHFB calculations. As a result, the enhancement of the binding energy from spherical $^{11}$Be to the prolate minimum with odd parity is not so remarkable; see Fig. \ref{Fig:Ebeta} (b). In contrast, the nearby $1d_{5/2}$ component can be easily enfolded into the even-parity orbit $1/2_2^+$, following the deformation from spherical $^{11}$Be to the GS. Thus, due to strong couplings of the orbitals $1d_{5/2}$ and $1s_{1/2}$ with the core orbits, more evident enhancement following the deformation is found in even-parity $^{11}$Be than that in odd-parity case, which is crucial for correctly determining the even-parity GS of $^{11}$Be.

\subsection{The halo structure of $^{11}$Be}

\begin{figure*}
\begin{center}
  \includegraphics[width=0.5\linewidth]{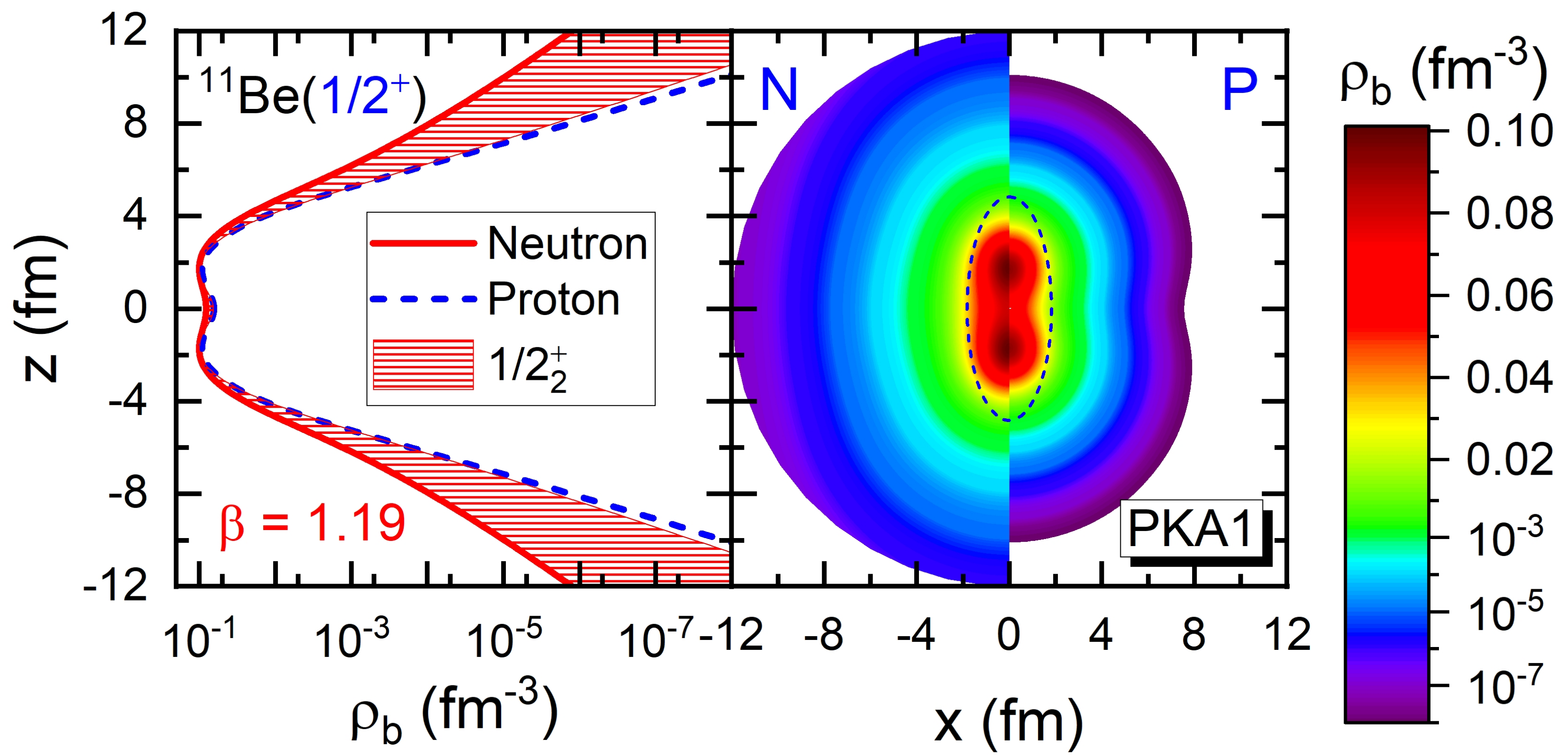}
  \figcaption{(color online) Neutron ($N$) and proton ($P$) densities (fm$^{-3}$) of $^{11}$Be given by PKA1, including logarithm ones (left plot) along the $z$-axis and contour maps (right plot) in which the dashed ellipsoid is defined by the deformation $\beta$ and radius $r$. }\label{Fig:DN-A1}
\end{center}
\end{figure*}

On the basis of correctly reproducing the parity of the GS, we further explore the halo phenomenon in $^{11}$Be. Figure \ref{Fig:DN-A1} shows the neutron ($N$) and proton ($P$) density profiles of the GS of $^{11}$Be given by PKA1, including the logarithm ones along the symmetric $z$-axis (left plot) and the contour maps (right plot), in which the dashed ellipsoid defined by the deformation $\beta$ and radius $r$ is shown as a reference for the size and shape of $^{11}$Be. As shown in Fig. \ref{Fig:DN-A1}, the neutrons in $^{11}$Be distributed rather extensively along the $z$-axis. Different from the decoupling picture \cite{Zhou2010PRC82.011301, Sun2018PLB785.530}, the neutron dispersion in $^{11}$Be shows a consistent shape with the whole nucleus. Moreover, it can be seen from the left plot of Fig. \ref{Fig:DN-A1} that the neutron dispersion is almost fully a results of the even-parity orbit $1/2_2^+$ (in pattern of horizontal lines), as referred to in the proton density profiles. It is worth noting that such results are consistent with the experimental indication of a one-neutron halo \cite{Fukuda1991PLB268.339, Aumann2000PRL84.35, Schmitt2012PRL108.192701, Tanihata2013PPNP68.215}.

However, the orbit $1/2_2^+$ in Fig. \ref{Fig:SPK-Be} (b) is somewhat deeply bound for the GS of $^{11}$Be, which seems to conflict with the extensive neutron distributions in Fig. \ref{Fig:DN-A1}. Combined with the results in Fig. \ref{Fig:SPK-Be}, a picture becomes apparent. As shown in Fig. \ref{Fig:SPK-Be} (b), the neutron orbit $2s_{1/2}$ is loosely bound for spherical $^{11}$Be because of the nodal difference, and for the same reason, the $s_{1/2}$ fragment (mainly the $2s_{1/2}$ one) remains loosely bound in the largely deformed GS. Due to the vanishing centrifugal barrier, this leads to the halo occurrence in the GS of $^{11}$Be. In contrast, as hindered by the centrifugal barrier, the neighboring $d_{5/2}$ fragment does not contribute to the halo, but plays an essential role in stabilizing the halo via the attractive inherent correlations with the $s_{1/2}$ fragment, as indicated by the negative $E_{c,i}$ value of the halo orbit $1/2_2^+$ aforementioned.

In fact, the enhancements due to the $\rho$-T and $\pi$-PV couplings are not only crucial for the parity inversion but also significant for the formation and stabilization of the halo in $^{11}$Be. Following the deformation from spherical $^{11}$Be to the GS, the spherical fragment $2s_{1/2}$ remains weakly bound to form the halo, and the intruding fragment $1d_{5/2}$ in the orbit $1/2_2^+$ enhances the couplings with the core orbits of $^{11}$Be, resulting in an even-parity GS, in which the $\rho$-T and $\pi$-PV couplings play an important role. This also stabilizes the halo via an attractive inherent correlation with the $2s_{1/2}$ fragment. Moreover, due to the nodal difference between the $2s_{1/2}$ and $1d_{5/2}$ components, such an inherent correlation is not strong enough to hinder halo formation. It is worth noting that this stability mechanism is different from that of halos in spherical unstable nuclei, which are stabilized in general via enhanced pairing correlations by neighboring high-$l$ orbits \cite{Meng2006PPNP57.470}.

\section{Summary}\label{sec:Summary}

In summary, employing the D-RHFB model, we explored the mechanism behind the parity inversion and neutron halo of $^{11}$Be, which is uniformly interpreted by the RHF Lagrangian PKA1. Using the language of the spherical DWS basis expansion, it is illustrated that evidently enhanced deformation effects by the $\rho$-T and $\pi$-PV couplings are crucial for describing correctly the even-parity ground state and the neutron halo of $^{11}$Be correctly. Moreover, a new microscopic picture of the halo formation and stability is revealed in this work, which paves a way to understanding the halo in deformed unstable nuclei. For light nuclei, the effects of the rotation and vibration corrections might be important. In the future, further implementations of the D-RHFB model are expected by considering the angular momentum projection, the number projection, and other beyond mean field effects.

\section{Acknowledgments}

\emph{The authors want to thank Prof. S.-G. Zhou, Prof. F. Q. Chen and Dr. X. X. Sun for fruitful discussions and the Supercomputing Center of Lanzhou University and the Southern Nuclear Science Computing Center for the computing resources provided.}

\end{multicols}
\vspace{10mm}
\begin{multicols}{2}

\end{multicols}

\end{document}